\documentclass[twocolumn,aps,prl,amsmath,floatfix]{revtex4}
\usepackage{graphicx}
\begin{document}
\title{High-energy pseudogap in degenerate Hubbard model induced via Hund coupling}
\author{Ansgar Liebsch} 
\affiliation{Peter Gr\"unberg Institute and Institute of Advanced Simulations, 
             Forschungszentrum J\"ulich, 52425 J\"ulich, Germany} 
\begin{abstract}
Hund coupling in the degenerate five-band Hubbard model near $n=6$ occupancy is shown 
to give rise to a significant depletion of spectral weight above the Fermi level. 
Calculations within dynamical mean field theory combined with exact diagonalization 
reveal that this pseudogap is associated with a collective mode in the self-energy 
caused by spin fluctuations. The pseudogap is remarkably stable over a wide range of 
Coulomb and exchange energies, but disappears for weak Hund coupling. The implications 
of this phenomenon for optical spectra of iron pnictides are discussed. 
\\
\mbox{\hskip1cm}  
electron systems 
\end{abstract}
\maketitle

The discovery of superconductivity in iron pnictides \cite{kamihara} 
has stimulated intense discussions concerning the role of correlation effects in 
these compounds. In contrast to high-$T_c$ cuprates, which have antiferromagnetic 
Mott insulators as  parent compounds, pnictides are correlated magnetic metals 
that tend to show significant deviations from Fermi-liquid behavior. Moreover, 
as a result of the multi-band nature of pnictides, the interplay of Coulomb 
and exchange interactions give rise to phenomena not found in cuprates.
The importance of Hund coupling in pnictides and chalcogenides was recently 
pointed out in several theoretical \cite{haule09,johannes,lauchli,zhou,yin} and 
experimental studies \cite{hu,wang,schafgans}. 
Optical data on paramagnetic LaFePO \cite{qazilbash} and BaFe$_2$As$_2$ 
\cite{hu,wang,schafgans} reveal a high-energy pseudogap not compatible with normal 
metal behavior. This pseudogap differs fundamentally from the low-energy gap 
in the antiferromagnetic spin-density wave phase.  Also, angle-resolved 
photoemission spectra for doped BaFe$_2$As$_2$ exhibit a depletion of spectral 
weight near the Fermi level that differs from the superconducting gap \cite{xu}.  

To investigate the influence of Coulomb correlations on the electronic 
properties of iron pnictides, various groups \cite{haule09,yin,haule08,craco,
aichhorn09,skornyakov,prb2010a,hansmann,aichhorn10,prb2010b,aichhorn11,werner11}
have used dynamical mean field theory \cite{dmft} (DMFT) combined with 
single-particle Hamiltonians derived from density-functional theory. These 
calculations typically revealed moderate to strong effective mass enhancement, 
in agreement with experimental data. Correlations were also shown to give rise 
to nonzero low-energy scattering rates \cite{craco,prb2010a,aichhorn10}, 
indicating bad metallicity. On the other hand, 
as a result of the complex band structure of pnictides, all five $d$ bands 
are important, so that the many-body features exhibit a marked orbital 
dependence. Moreover, the Fe $3d$ density of states generally shows two 
main features separated by a deep minimum slightly above the Fermi energy.
In view of these multi-band characteristics, it is difficult to distinguish 
genuine many-body features from single-particle properties. For instance, it 
is presently not clear to what extent pseudogaps in the interacting density 
of states are induced by Coulomb correlations or band structure effects.    
   
The aim of the present work is to unravel these competing influences. 
For this purpose we have performed DMFT calculations for a simplified 
Hamiltonian consisting of five degenerate semi-elliptical bands. 
Since the density of states is featureless, correlation induced spectral 
modifications are easily identified. In addition, by scanning a wide 
range of Coulomb and exchange energies, it is feasible to investigate their 
respective roles. The main result of this work is that many-body effects are 
much more sensitive to the magnitude of Hund's rule coupling $J$ than to the 
intra-orbital Coulomb repulsion $U$. In particular, exchange interactions 
give rise to a pseudogap above $E_F$, which persists in the full range of 
realistic values of $J$, but disappears at small $J$. Within the present 
model, this pseudogap can be linked to a resonance in the self-energy caused
by spin fluctuations. We also argue that these correlation effects can be 
understood qualitatively by viewing the system at $n=6$ occupancy as a doped 
$n=5$ Mott insulator. 
Because of the high orbital degeneracy, the paramagnetic half-filled system is 
metallic at small $J$, but insulating already at moderate Hund coupling. 
Upon electron doping, a narrow quasi-particle peak appears in the density 
of states, with a maximum below and a pseudogap above $E_F$. Thus, as a 
function of $J$, the pseudogap at $n=6$ exists only in proximity to the 
Mott phase of the half-filled system.

To motivate the present work, we show first in Fig.~1 a comparison of the 
electron spectral distribution of FeAsLaO derived within DMFT with the bare 
density of states obtained from the effective $3d$ single-particle Hamiltonian 
provided in Ref.~\onlinecite{miyake}. Both spectra exhibit a 
striking minimum slightly above $E_F$. As impurity solver we use exact 
diagonalization (ED) \cite{ed}. Each $d$ orbital hybridizes with one or two 
bath levels, giving total cluster sizes $n_s=10$ or $n_s=15$, respectively. 
The interacting Hamiltonian and additional details concerning the ED DMFT approach 
are specified in Ref.~\onlinecite{prb2010a}. Coulomb and exchange interactions 
are treated in a rotationally invariant manner. The temperature 
of the Matsubara grid corresponds to $T\approx 0.01$~eV, while the interacting 
cluster Green's function is evaluated at $T=0$ for computational reasons. 
The spectral distribution at real $\omega$ is obtained by using the routine 
{\it ratint} \cite{ratint} to extrapolate the lattice Green's function 
from the imaginary axis. Typically, a few hundred Matsubara points are used 
in this extrapolation, with a frequency dependent broadening. 
This procedure yields reliable results at low frequencies. Uncertainties 
arise mainly at higher frequencies and affect the position and width of the 
Hubbard bands. 

\begin{figure} [t!] %1
\begin{center}
\includegraphics[width=4.5cm,height=6.5cm,angle=-90]{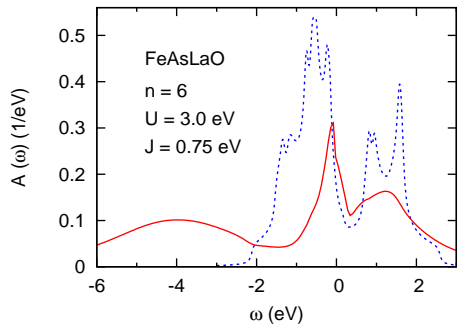}  %{A.FeAs.ps}
\end{center}\vskip-4mm
\caption{(Color online)
Fe $3d$ quasiparticle spectra of FeAsLaO, calculated within 
ED DMFT ($n_s=15$) for $U=3$~eV, $J=0.75$~eV (solid red curve).
The bare $3d$ density of states is indicated by the dashed blue curve
\cite{miyake}.
}\end{figure}

The quasiparticle spectrum of FeAsLaO is seen to exhibit two main peaks, 
below and above $E_F$, separated by a narrow pseudogap slightly above 
$E_F$, and a lower Hubbard band below the Fe $3d$ band region. The bare 
density of states also exhibits pronounced peaks below and above $E_F$, 
and a deep minimum about $0.3$~eV above the Fermi energy. Comparing these 
spectra one is tempted to conclude that Coulomb interactions lead to a 
significant narrowing of the occupied $3d$ bands and to a strong shift 
towards $E_F$. Only a small fraction of the occupied spectral weight 
($\sim 0.15/0.6$ per spin band) remains in the coherent peak near $E_F$, 
while the main part ($\sim 0.45/0.6$) is distributed over the wide incoherent 
region. In the unoccupied region, the effect of Coulomb interactions is less 
severe. The main density of states peak is merely broadened and the minimum 
about $0.3$~eV above $E_F$ is approximately preserved. Similar quasi-particle 
spectra have been obtained for a variety of pnictides \cite{yin,craco,aichhorn09,
prb2010a,hansmann,skornyakov,prb2010b,aichhorn10,aichhorn11,werner11,ikeda,ising}. 
In principle, integer occupancy $n=6$ can also sustain a Mott phase. This phase 
occurs, however, at much larger Coulomb energies \cite{lauchli,prb2010a,aichhorn10}.     

\begin{figure} [t!] %2
\begin{center}
\includegraphics[width=4.5cm,height=6.5cm,angle=-90]{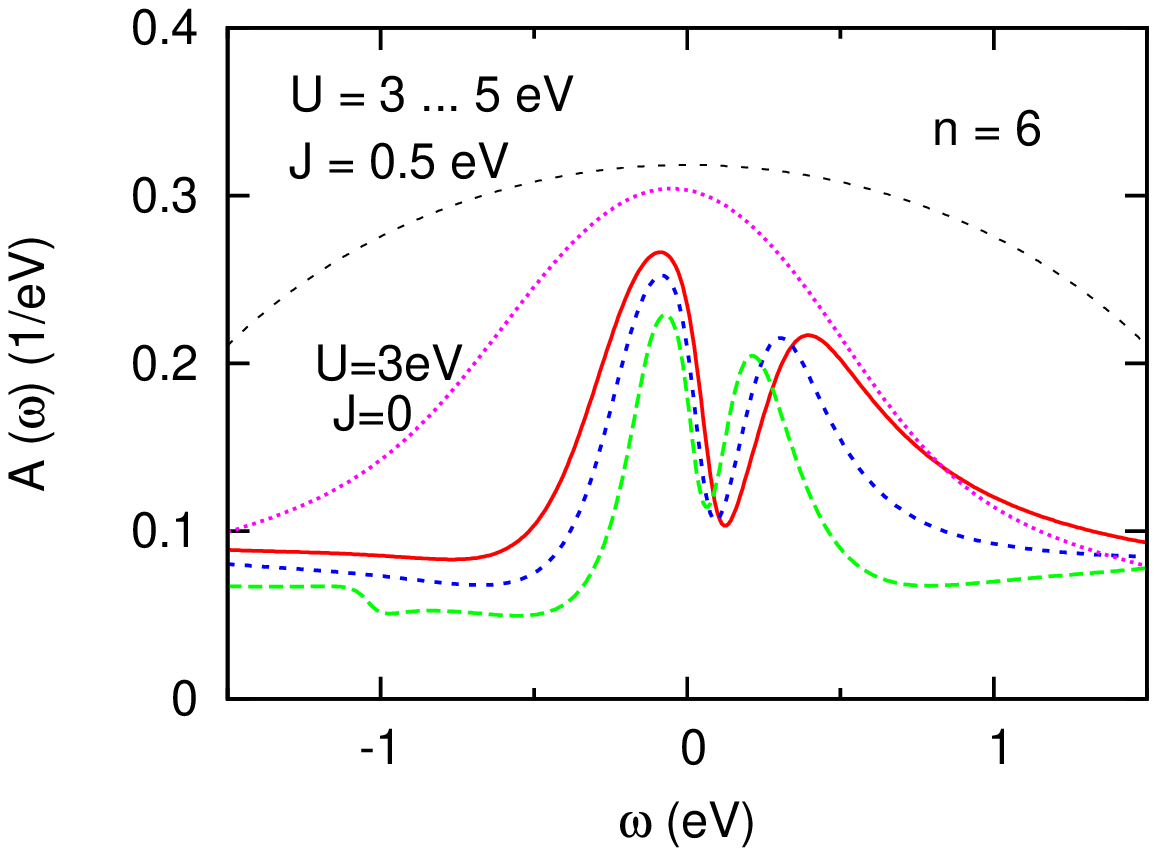} %{Gnugr1.ps}
\end{center}
\vskip-7mm \ \ \ (a)\hfill  \mbox{\hskip5mm}
\begin{center}
\vskip-5mm
\includegraphics[width=4.5cm,height=6.5cm,angle=-90]{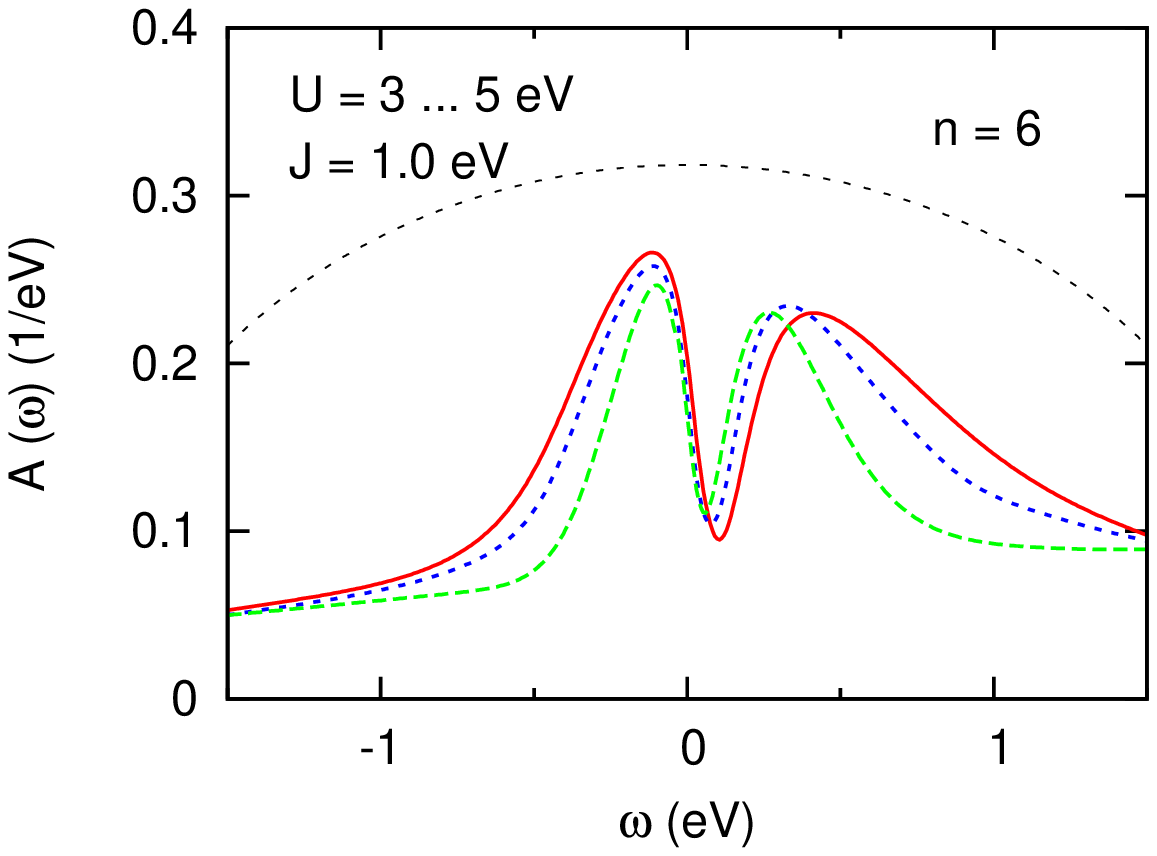}  %{Gnugr2.ps}
\end{center}
\vskip-7mm \ \ \ (b)\hfill  \mbox{\hskip5mm}\vskip-1mm
\caption{(Color online) 
Low-energy region of spectral distribution of degenerate five-band model at 
occupancy $n=6$, calculated within ED DMFT ($n_s=10$) for several Coulomb 
energies: $U=3$~eV (solid red curve), $U=4$~eV (short-dashed blue curve), 
and $U=5$~eV (long-dashed green curve). (a) $J=0.5$~eV, (b) $J=1.0$~eV. 
Spectra at intermediate values of $J$ are similar. 
The upper panel also shows the results for $U=3$~eV, $J=0$.
The bare density of states is indicated by the dotted curve.
}\end{figure}

Although the scenario discussed above appears plausible, it is nontrivial to 
disentangle genuine many-body features from the complex single-particle aspects 
of the density of states. In order to identify interaction effects due to 
$U$ and $J$ we have carried out ED DMFT calculations for a Hamiltonian 
consisting of five identical subbands with semi-elliptical density of states, 
where the width $W=4$~eV corresponds to typical pnictide compounds.  
Fig.~2 shows spectral distributions 
for a wide range of Coulomb and exchange energies, $U=3,\ldots,5$~eV and 
$J=0.5,\ldots,1.0$~eV. Most previous DMFT studies of pnictides used $U$ 
and $J$ values within this range. For simplicity, the present calculations  
are performed using one bath level per $3d$ orbital ($n_s=10$). 
The results agree qualitatively with more accurate ones for two bath levels 
per orbital. Even though $U$ and $J$ span a wide range, the spectra are seen 
to be remarkably similar. In all cases, the interacting density of states 
exhibits peaks below and above $E_F$, separated by a pseudogap slightly 
above $E_F$. The peak--dip structure near $E_F$ is reminiscent of the one 
in the quasi-particle spectrum shown in Fig.~1. 
Since in the present model the bare density of states is featureless,  
the pseudogap is entirely due to the frequency-dependent self-energy. 
Moreover, the pseudogap depends only weakly on the values of $U$ and $J$ 
within the range quoted above, but it disappears at small $J$, so that only 
a broad quasiparticle peak remains. An example is shown in Fig.~2(a) for 
$U=3$~eV, $J=0$. These results suggest that the pseudogap above $E_F$ is a 
generic feature caused by multiband Coulomb correlations within the $3d$ shell 
and that its existence depends crucially on realistic values of Hund coupling. 
The paramagnetic quasi-particle distribution of actual pnictides should 
therefore consist of a combination of correlation features associated with 
$J$ and signatures related to the bare density of states.

\begin{figure} [t!] %3
\begin{center}
\includegraphics[width=4.5cm,height=6.5cm,angle=-90]{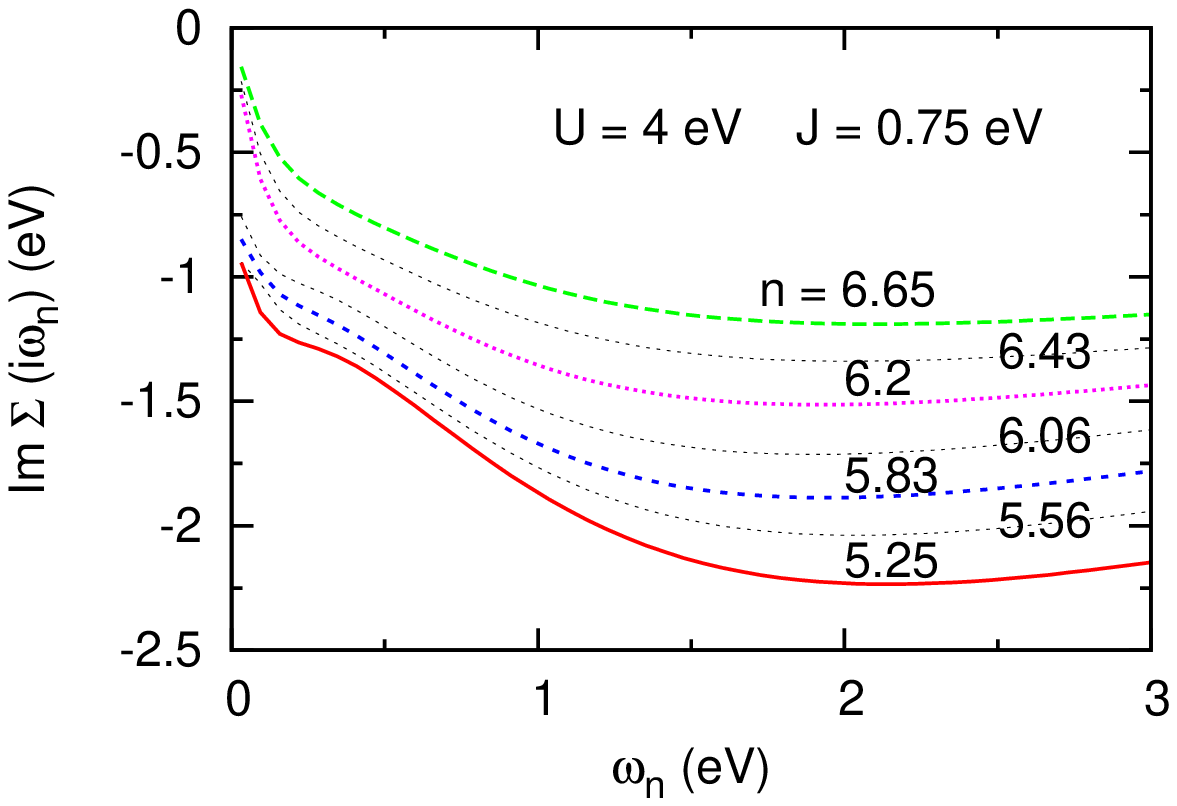}  %{Sigmai.ps}
\end{center}
\vskip-7mm \ \ \ (a)\hfill  \mbox{\hskip5mm}
\begin{center}
\vskip-5mm
\includegraphics[width=4.5cm,height=3.9cm,angle=-90]{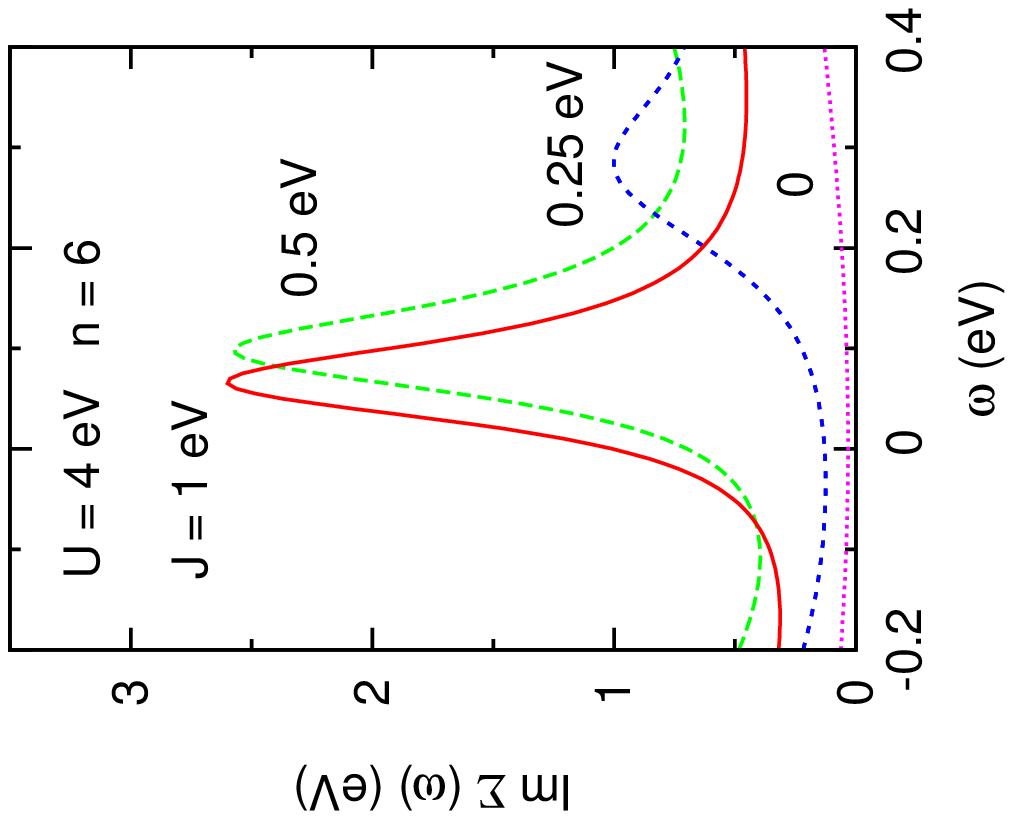}  %{Gnusr.ps}
\includegraphics[width=4.5cm,height=3.9cm,angle=-90]{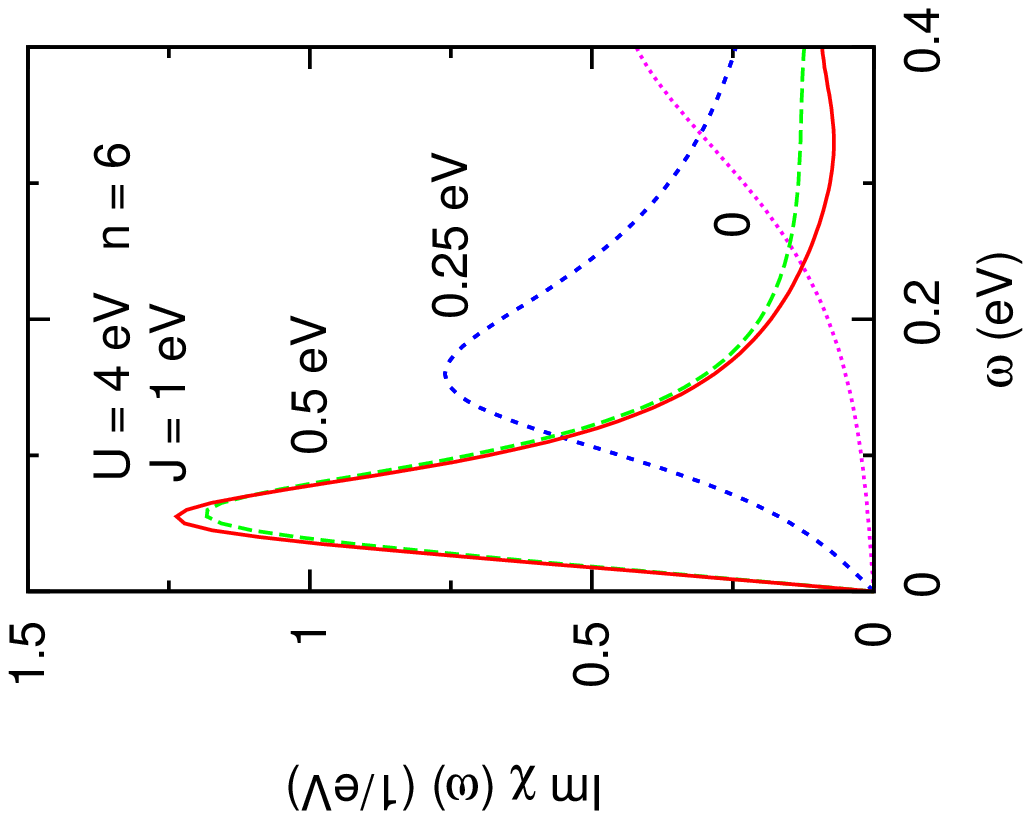}  %{Gnucr.ps}
\end{center}
\vskip-7mm \ \ \ (b)\hfill  \mbox{\hskip5mm}\vskip-1mm
\caption{(Color online)
(a) Imaginary part of self-energy for degenerate five-band model, calculated 
within ED DMFT ($n_s=15$) as a function of Matsubara frequency for various 
occupancies at $U=4$~eV,  $J=0.75$~eV. (b) Variation of self-energy (left) 
and spin susceptibility (right) with Hund coupling for $U=4$~eV, $n=6$ ($n_s=10$). 
}\end{figure}

To illustrate the effect of Coulomb correlations on the self-energy, we show 
in Fig.~3 the imaginary part of $\Sigma(i\omega_n)$ for several occupancies. 
For $n\ge 6.2$, Im\,$\Sigma(i\omega_n)\sim\omega_n$ at low energies, so that 
the system is a correlated Fermi liquid with a doping dependent effective mass 
enhancement $m^*/m\approx 3,\ldots,6$. At smaller occupancy, Im\,$\Sigma$ 
reveals a nonzero onset, leading to bad-metallic behavior. As shown previously 
\cite{prb2010b}, the spin correlation function then changes from Pauli to Curie 
Weiss behavior, indicating a spin freezing transition, in close analogy to the  
one found in the degenerate three-band model near $n=2$~\cite{werner}. 
In addition, Im\,$\Sigma(i\omega_n)$ exhibits a novel kink near 
$\omega_n\approx 0.1,\ldots,0.2$~eV, which is weak in the Fermi-liquid region, 
but becomes more intense at lower occupancy. As shown in panel (b), this kink 
is related to a resonance in Im\,$\Sigma(\omega)$, which is the origin of the 
pseudogap above $E_F$ in the spectra shown in Fig.~2. Remarkably, the resonance 
disappears at small $J$. For $J=0.5, \ldots,1.0$~eV, the local spin correlation 
function reveals a maximum at about $0.1$~eV, suggesting that the resonance in 
Im\,$\Sigma(\omega)$ corresponds to a collective mode induced by spin fluctuations
associated with Hund coupling. Thus, states in the resonance  
region have a greatly reduced lifetime. According to Kramers-Kronig relations, 
Re\,$\Sigma(\omega)$ exhibits a positive slope near the resonance, so that 
spectral weight is removed from the pseudogap region. Beyond this region, the 
slope of Re\,$\Sigma(\omega)$ becomes again negative, giving rise to a kink 
in the dispersion of energy bands \cite{byczuk}.

\begin{figure} [t!] %4
\begin{center}
\includegraphics[width=4.5cm,height=6.5cm,angle=-90]{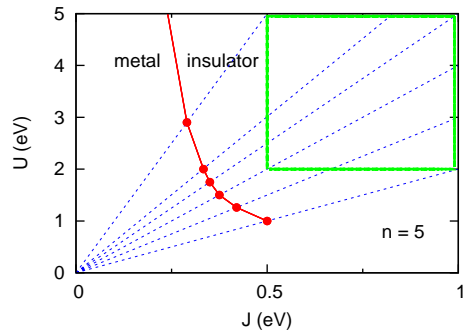} %{phase.ps}
\end{center}
\vskip-4mm 
\caption{(Color online) 
Paramagnetic phase diagram of degenerate five-band model at half-filling.
The dashed blue lines denote fixed ratios $J/U$ between $1/10$ and $1/2$. 
The dots represent the critical Coulomb energies of the Mott transitions
along these lines. To the right (left) of the solid red curve, the system 
is insulating (metallic). The green box indicates the range of $U$ and $J$ 
values commonly used in DMFT studies of iron pnictides. For $n=6$ the Mott 
phase occurs at much larger values of $U$ \cite{lauchli}, so that, 
approximately to the right of the solid red curve, the system is bad metallic.
}\end{figure}

An alternative, complementary understanding of the pseudogap can be achieved by 
realizing that the system at $n=6$ occupancy exists in proximity to the $n=5$ 
Mott insulator. As shown in Fig.~4, in the limit of half-filling, the degenerate 
five-band model exhibits a Mott transition at a critical Coulomb energy that depends 
very sensitively on $J$, but is rather insensitive to the value of $U$. 
In particular, $U_c$ diminishes rapidly when $J$ increases: $U_c\approx3$~eV for 
$J=0.25$~eV and $U_c\approx1$~eV for $J=0.5$~eV. (For $J=U/4$ and $J=U/6$, we 
find $U_c\approx1.5$~eV and $U_c\approx2.0$~eV, respectively, in agreement with 
Ref.~\onlinecite{lauchli}.) In contrast, because of the high orbital degeneracy, 
$J=0$ yields a Mott gap only at very large Coulomb energy ($U_c\approx 15$~eV). 
Thus, for $U$ and $J$ values usually employed in  DMFT studies of 
iron pnictides (see green box), the half-filled system is a Mott insulator 
composed of localized $S=5/2$ spins.   
Upon electron doping, some of the localized states become itinerant, so 
that the density of states exhibits a narrow quasiparticle peak below $E_F$, 
and a concomitant minimum or pseudogap above $E_F$, in agreement with the 
spectra shown in Fig.~2. DMFT calculations as a function of doping 
(not shown here) reveal that, away from half-filling, the upper Hubbard band 
quickly gets weaker and merges with the spectral feature above $E_F$, while the 
lower Hubbard peak remains rather stable. On the other hand, according to the 
phase diagram shown in Fig.~4, in the absence of exchange interactions, the 
high orbital degeneracy yields a weakly correlated metal since $U\ll U_c$. 
This also holds at finite electron doping, so that the spectral distribution
exhibits a broad quasi-particle peak, as indicated by the results for $U=3$~eV, 
$J=0$ in Fig.~2(a).  It is evident, therefore, that the collective mode in the 
self-energy shown in Fig.~3 and the associated pseudogap in the density of states 
(Fig.~2) obtained at realistic values of $J$ are a consequence of Hund coupling. 

As pointed out above, the complex geometry and band structure of pnictides 
gives rise to appreciable orbital dependence of the self-energy and interacting 
density of states, as has been found in various DMFT studies 
\cite{yin,haule08,craco,aichhorn09,skornyakov,prb2010a,hansmann,aichhorn10}. 
In addition, screening processes differ greatly among pnictide compounds, 
leading to substantial variation in Coulomb and exchange energies \cite{miyake}.     
Resonances in Im\,$\Sigma(\omega)$ and the corresponding pseudogaps in 
the interacting density of states might appear therefore only in certain $3d$ 
orbitals. This seems to be the case, as indicated, for instance, by the kinks 
found in some Im\,$\Sigma(i\omega_n)$ components of FeAsLaO \cite{prb2010a}. 
Nonetheless, the robustness of the pseudogap in Fig.~2 over a wide range of 
$U$ and $J$ suggests that the characteristic peak--dip structure near $E_F$ 
seen in quasiparticle spectra of pnictides \cite{yin,craco,aichhorn09,prb2010a,
hansmann,skornyakov,prb2010b,aichhorn10,aichhorn11,werner11,ikeda}
might be partially due to Coulomb correlations. 
Optical spectra in the paramagnetic phase then should exhibit a depletion 
of spectral weight close to $E_F$, as was recently predicted theoretically 
\cite{yin} and observed experimentally in BaFe$_2$Se$_2$ \cite{hu,wang,schafgans}. 
In the future, it would be very interesting to search more systematically for 
possible links between pseudogaps and collective modes in the self-energy 
of pnictides, and to study their variation with doping and temperature.

The picture discussed above has more general validity beyond the present five-band model. 
Thus, pseudogaps generated by self-energy resonances due to Hund coupling are 
also found in the analogous three-band system \cite{tobe}. The main difference 
is that, because of the lower degeneracy, the boundary in Fig.~4 between metallic 
and insulating phases at half-filling is shifted to larger $J$ (except at large $U$)
\cite{werner79} and that at $n=3\pm1$ the system is relatively farther 
from the Mott phase than at $n=5\pm1$.
 
In summary, Hund coupling in ferropnictides plays an important role in the 
formation of a high-energy pseudogap. To disentangle this gap from density 
of states effects, the non-interacting part of the Hamiltonian is simplified 
in terms of five degenerate semi-elliptical subbands. The pseudogap can then 
be identified as a many-body feature that is stable over a wide range of 
Coulomb and exchange energies and that is related to a collective mode in the 
self-energy associated with spin fluctuations. 
The nature of this mode can also be understood by viewing the system as 
a doped $n=5$ Mott insulator, where itinerant electrons coexist with localized 
spins. This picture is consistent with the observation that at finite doping 
the spectral distribution exhibits a narrow quasiparticle peak below $E_F$ and 
a concomitant minimum above $E_F$, while most of the spectral weight resides 
in the lower Hubbard band. The crucial role of Hund exchange in this scenario 
is evident from the fact that, due to the high orbital degeneracy, at small 
values of $J$ the pseudogap disappears and the system is a weakly correlated 
metal. Thus, the high-energy pseudogap is directly linked to the realistic 
magnitude of Hund's rule coupling.

I like to thank Hiroshi Ishida for fruitful discussions.
The DMFT calculations were carried out on the J\"ulich Juropa machine.


\begin{thebibliography}{99} %xxxxxxxxxxxxx

\bibitem{kamihara}
   Y. Kamihara, T. Watanabe, M. Hirano, and H. Hosono, 
   J. Am. Chem. Soc. {\bf 130}, 3296 (2008).

\bibitem{haule09}
   K. Haule and G. Kotliar, 
   New. J. Phys. {\bf 11}, 025021 (2009).

\bibitem{johannes}
   M. D. Johannes and I. I. Mazin,
   Phys. Rev. B  {\bf 79}, 220510(R) (2009).

\bibitem{lauchli}
   A. M. L\"auchli and Ph. Werner,
   Phys. Rev. B  {\bf 80}, 235117    (2009).

\bibitem{zhou}
   S. Zhou and Z. Wang,
   Phys. Rev. Lett. {\bf 105}, 096401 (2010).

\bibitem{yin}
   Z. P. Yin, K. Haule, and G. Kotliar,
   Nature Physics {\bf 7}, 294 (2011).

\bibitem{hu}
   W. Z. Hu {\it et al.}, % , J. Dong, G. Li, Z. Li, P. Zheng,  G. F. Chen, J. L. Luo, and N. L. Wang,
   Phys. Rev. Lett. {\bf 101}, 257005 (2008).

\bibitem{wang}
   N. L. Wang {\it et al.}, % , W. Z. Hu, Z. G. Chen, R. H. Yuan, G. Li, G. F. Chen, and T. Xiang,
   arXiv:1105.3939.

\bibitem{schafgans}
   A. A. Schafgans {\it et al.}, % , S. J. Moon, B. C. Pursley, A. D. LaForge, M. M. Qazilbash,   A. S. Sefat, D. Mandrus, K. Haule, G. Kotliar, and D. N. Basov, 
   arXiv:1106.3114.

\bibitem{qazilbash}
   M. M. Qazilbash {\it et al.}, % , J. J. Hamlin, R. E. Baumbach, Lijun Zhang, D. J. Singh,  M. B. Maple, and D. N. Basov,
   Nature Physics {\bf 5}, 647 (2009). 

\bibitem{xu}
   M. Xu {\it et al.}, % , P. Richard, K. Nakayama, T. Kawahara, Y. Sekiba, T. Qian, M. Neupane, S. Souma, T. Sato, T. Takahashi, H.-Q. Luo, H.-H. Wen, G.-F. Chen, N.-L. Wang, Z. Wang, Z. Fang, X. Dai, and  H. Ding, 
   Nature Communications {\bf 2}, 392 (2011).

%%%%%%%%%%%%%%%%%%%%%%%%%%%%%%%%%%%%%%%%%%%%%%%%%%%%%%%%%%%%%%%%%%


\bibitem{haule08}
   K. Haule, J. H. Shim, and G. Kotliar, 
   Phys. Rev. Lett. {\bf 100}, 226402 (2008).

\bibitem{craco}
   L. Craco {\it et al.}, % ,  M. S. Laad, S. Leoni, and H. Rosner,
   Phys. Rev. B {\bf 78}, 134511 (2008).
%  M. S. Laad, L. Craco, S. Leoni, and H. Rosner,
%  Phys. Rev. B {\bf 79}, 024515 (2009).
%   L. Craco, M. S. Laad, and S. Leoni, arXiv:0910.3828
%   (unpublished).

\bibitem{aichhorn09}
   M. Aichhorn {\it et al.}, % , L. Pourovskii, V. Vildosola, M. Ferrero, O. Parcollet, T. Miyake, A. Georges, and S. Biermann,  
   Phys. Rev. B {\bf 80}, 085101 (2009).

\bibitem{skornyakov}
   S. L. Skornyakov {\it et al.}, % N. A. Skorikov, A. V. Lukoyanov, A. O. Shorikov, and V. I. Anisimov,
   Phys. Rev. B {\bf 80}, 092501 (2009); {\it ibid.} {\bf 81}, 174522 (2010). 

\bibitem{prb2010a}
  H. Ishida and A. Liebsch, 
  Phys. Rev. B {\bf 81}, 054513 (2010).

\bibitem{hansmann} 
    P. Hansmann {\it et al.}, % , R. Arita, A. Toschi, S. Sakai, G. Sangiovanni, and K. Held,
    Phys. Rev. Lett. {\bf 104}, 197002 (2010).

\bibitem{aichhorn10}
   M. Aichhorn {\it et al.}, % , S. Biermann, T. Miyake, A. Georges, and M. Imada,
   Phys. Rev. B {\bf 82}, 064504 (2010).

\bibitem{prb2010b}
   A. Liebsch and H. Ishida, 
   Phys. Rev. B {\bf 82}, 155106 (2010).

\bibitem{aichhorn11}
   M. Aichhorn {\it et al.}, % L. Pourovskii, and A. Georges,
   arXiv:1104.4361.
 
\bibitem{werner11}
   Ph. Werner {\it et al.}, % , M. Casula, T. Miyake, F. Aryasetiawan, A. J. Millis, and S. Biermann,
   arXiv:1107.3128.

\bibitem{dmft}
   A. Georges {\it et al.}, % , G. Kotliar, W. Krauth, and M. J. Rozenberg, 
   Rev. Mod. Phys. {\bf 68}, 13 (1996).


%%%%%%%%%%%%%%%%%%%%%%%%%%%%%%%%%%%



\bibitem{miyake}
   T. Miyake {\it et al.}, % , K. Nakamura, R. Arita, and M. Imada,
   J. Phys. Soc. Jpn. {\bf 79}, 044705 (2010).

\bibitem{ed}  
   M. Caffarel and W. Krauth, 
      Phys. Rev. Lett. {\bf 72}, 1545 (1994).

\bibitem{ratint}
   {\it Numerical Recipes in Fortran 77},
   Cambridge University Press, p. 106 (1986-1992).

\bibitem{ikeda}
   H. Ikeda, R. Arita, and J. Kunes,
   Phys. Rev. B {\bf 82}, 024508 (2010).
   
\bibitem{ising}
   Density-density exchange treatments amount to an enhancement of Coulomb 
   interactions \cite{prb2010a} and can lead to dramatic redistribution of 
   spectral weight near $E_F$ \cite{aichhorn10}.  

\bibitem{werner}
   P. Werner {\it et al.}, % , E. Gull, M. Troyer, and A. J. Millis,
   Phys. Rev. Lett.  {\bf 101}, 166405  (2008).

\bibitem{byczuk}
   For correlation-induced kinks in one-band models, and their relation to collective modes
   due to spin fluctuations, see:
   K. Byczuk {\it et al.}, % , M. Kollar, K. Held, Y.-F. Yang, I. A. Nekrasov, T. Pruschke,  and D. Vollhardt, 
   Nature Physics {\bf 3}, 168 (2007);
   C. Raas {\it et al.},   Phys. Rev. Lett.  {\bf 102}, 076406  (2009);
   P. Grete {\it et al.},  arXiv:1107.1370.

\bibitem{tobe}
   A. Liebsch, unpublished.

\bibitem{werner79}
   Ph. Werner, E. Gull, and A. Millis, 
   Phys. Rev. B  {\bf 79}, 115119    (2009).

%\bibitem{luca}
%   L. de' Medici {\it et al.}, %, J. Mravlje, and A. Georges,
%   arXiv:1106.0815.
   

\end{thebibliography}
\end{document}